\documentclass{kluwer}    % Specifies the document style.
\usepackage{graphicx}
\newdisplay{guess}{Conjecture}

\begin{document}                                                                                   
\begin{article}
\begin{opening}         
\title{Updated Toulouse solar models including the
diffusion-circulation coupling and the effect of $\mu$-gradients} 
\author{Olivier \surname{Richard}}
\institute{GRAAL UMR5024, Universit\'e Montpellier II, CC072, Place E. Bataillon, 34095 Montpellier, France}
\author{Sylvie \surname{Th\'eado}}
\institute{Centro de Astrofisica da Universidade do Porto, Rua das Estrelas, 4150-762 Porto, Portugal}
\author{Sylvie \surname{Vauclair}}  
\institute{Laboratoire d'Astrophysique, 14 av. Edouard Belin,31400 Toulouse, France}
\runningauthor{Richard, Th\'eado and Vauclair}
\runningtitle{Updated Toulouse solar models}
\date{November 27, 2003}

\begin{abstract}
New solar models are presented, which have been computed with the most recent physical inputs (nuclear reaction rates, equation of state, opacities, microscopic diffusion). Rotation-induced mixing has been introduced in a way which includes the feed-back effect of the $\mu$-gradient induced by helium settling. A parametrization of the tachocline region below the convective zone has also been added in the computations. The sound velocities have been computed in the models and compared to the seismic sun. Our best model is described in some detail. Besides the new physical inputs, the most important improvement concerns the computations of $\mu$-gradients during the solar evolution and their influence in slowing down rotation-induced mixing. This process can explain why lithium is depleted in the present Sun while beryllium is not, and meanwhile why $^3$He has not increased at the solar surface for at least 3 Gyrs.
\end{abstract}

\end{opening}           

\section{Introduction}  

For a long time the effects of microscopic diffusion 
were neglected in the construction of solar models. During
the last two decades helioseismology has strongly constrained
our view of the Sun and the importance of microscopic diffusion
has been demonstrated by many authors (e.g. Bahcall et al., 1995; Christensen-Dalsgaard et al.,
 1996; Richard et al., 1996
(RVCD); Brun et al., 1998a; Turcotte et al., 1998; Schlattl, 2002; Schlattl and Salaris, 2003). 
Microscopic diffusion is now systematically introduced in standard models.

However some significant differences do remain between the helioseismic 
and theoretical standard models in the region below the 
convective zone. Several solutions for this discrepancy have been
tested, such as opacity changes (Tripathy and Christensen-Dalsgaard, 1998) or turbulent 
mixing (RVCD; Brun et al., 1998, 1999 and 2002). 
Including mixing below the convective zone is also necessary
to account for the lithium depletion observed at the solar surface.
This mixing however must not extend too deep as beryllium is not depleted
(Balachandran and Bell, 1997) and the $^3$He value has not increased significantly during
the last 3 Gyrs (Geiss and Glocker, 1998)
In previous papers, rotation-induced mixing was introduced in the
computations but the feed-back effect of $\mu$-gradients on the
mixing was neglected or extremely simplified. In RVCD, all 
macroscopic motions were
supposed to disappear for a given critical $\mu$-gradient,
which was adjusted to satisfy the chemical constraints.
Recently, the effects of diffusion-induced $\mu$-gradients
on the meridional circulation and their consequences have
been more deeply studied (Vauclair, 1999; Vauclair 
and Th\'eado, 2003; Th\'eado and Vauclair, 2003 a and b). 
Numerical tools have been built which can be introduced
in the computations of stellar evolution and applied 
to the solar case.

We have computed new models with the 
Toulouse-Geneva
Evolutionary Code, including the most recent physics 
for the opacities, equation of state, nuclear reaction
rates, etc.
We have also introduced diffusion and 
rotation-induced mixing
in a consistent way, including the effects of 
$\mu$-gradients. 

In section 2, we discuss the observed solar
parameters and their uncertainties; the new physical inputs are
presented in section 3 ; section 4 is devoted to
the constraints on the mixing processes involved
(rotation-induced mixing and tachocline); the
mixing theory is reviewed in section 5 and the
results, with our best model, are presented
in section 6.

\section{The observed parameters}
\label{sec:const}

The solar luminosity $L_{\odot}$, radius $R_{\odot}$, and age
$t_{\odot}$ allow one to constrain the initial Helium abundance, $Y_0$, 
and the mixing length parameter, $\alpha$, of the standard model.
The luminosity $L_{\odot}$ is obtained from the solar constant
measured by satellite and varies with the solar activity.
We take $3.8515 \times 10^{33}$ ergs s$^{-1}$
(Guenther et al., 1992), which is the mean value between the results of the
two space 
missions ERS on {\it Nimbus} 7 and ACRIM on {\it SMM}.
The slightly greater value used by
Bahcall \& Pinsonneault (1992) and the smaller value used by 
Bahcall et al. (1995)
lie inside the error bars (see Table 1).

Some recent papers have shown that the solar radius value of $(6.9599
\pm 0.0007) \times 10^{10}$ cm (Allen, 1976) was overestimated.
Two different techniques now lead to a smaller value for the radius. 
The first one uses solar f-modes:
Antia (1998) and Schou et al. (1997) have respectively obtained
$6.9578 \times 10^{10}$ cm and $(6.9568 \pm 0.0003)
\times 10^{10}$ cm.
The second technique uses the observation of the angular diameter of
the Sun and 
atmosphere models. Brown \& Christensen-Dalsgaard (1998) have found a smaller radius
of $(6.95508 \pm 0.00026) \times 10^{10}$ cm.
For the calibration of our best model we 
take the mean observed value $(6.95749 \pm 0.00241) \times 10^{10}$
cm, which is very close to the helioseismic value obtained with the f-modes.

The last parameter used in the calibration is the solar age.
Few determinations of the solar age have been done. The meteoritic
determination obtained by Guenther et al. (1992) and by Wasserburg
(See Appendix of Bahcall et al., 1995) is respectively $4.52 \pm 0.04$
Gyr and $4.57 \pm 
0.02$. Using the helioseismic frequencies Guenther \& Demarque (1997) find $4.5 \pm
0.1$ Gyr and Dziembowski et al. (1999) obtain $4.66 \pm 0.11$ Gyr.

The precise radius at the bottom of the convective zone is also given by the
seismic sound speed profile. Christensen-Dalsgaard et al. (1991) found $0.713 \pm
0.003$. This value was confirmed by Basu \& Antia (1997) with a smaller
uncertainty; they gave : $0.713 \pm 0.001$, when they used only their
models with a smooth composition profile at the bottom of the
convective zone.

The helioseismic inversion can also give the helium abundance in the
  convective zone (Dziembowski et al., 1991; Kosovichev, 1995; 
Basu and Antia, 1997; Richard et al., 1998).
The most important uncertainty on the determination comes from the
  equation of state. 
With the OPAL equation of state
(Rogers et al., 1996) the typical value obtained for the helium mass
  fraction $Y$ is $0.249 \pm 0.002$; with the MHD equation of state
  (D\"appen et al., 1998; Hummer and Mihalas, 1988; Mihalas et al., 1988) 
the results differ more from
  author to
  author ($0.246$ for Basu and Antia, 1997, $0.242$ for
  Richard et al., 1998) due to the different inversion method.
  Basu et al. (1999) have recently studied the
  equation of state on the upper convective zone. They have found better
  agreement for the MHD equation of state than for the OPAL one on the
  helium ionization zone. However more detailed studies need to be done
  to reduce these uncertainties.

The photospheric abundances of heavier elements used as constraints on our solar models are those determined by Grevesse \& Noels (1993).

The values used to calibrate our models are summarized in Table \ref{soltab1}.

\section{New physical inputs}  

The present models have been computed including
the most recent improvements in the physical inputs :

\begin{itemize}
\item  Equation of state :
Considering the high precision presently achieved 
in helioseismology, it becomes necessary to take into account 
the relativistic effect on the
electrons in the core of the solar model (Elliot and Kosovichev, 1998).
In the present computations, we have introduced the new OPAL equation of state
(Rogers and Nayfonov, 2002) which takes this effect into account.
\item  Opacities :
We use the latest OPAL opacities (Iglesias \& Rogers, 1996) completed by the
Alexander \& Ferguson (1994) low temperature opacities.
\item  Nuclear reaction rates :
We have updated the nuclear routines using the NACRE compilation
of nuclear reaction rates (Angulo et al., 1999) and added Bahcall's 
screening routine.
\item  Microscopic diffusion : we have improved the computation of the 
microscopic
diffusion velocity by using equation 4.7 of Montmerle \& Michaud (1976) 
for all elements
other than helium. Turcotte et al. (1998) have pointed out a difference in 
the surface
abundance variations of C, O, and other metals (Z), during the evolution, between the 
Toulouse and
Montreal solar models (see table 3 of Turcotte et al., 1998). This 
difference was mainly due
to the thermal part of the diffusion velocity which was underestimated 
in the expression
used by RVCD. With the Montmerle \& Michaud
expression, the Toulouse
models have very similar surface abundance variations of C, O, and Z as 
those of the
Montreal models
\end{itemize}

\section{Constraints on mixing processes}

\subsection{Chemical constraints on mixing processes}

Models including helium and metal diffusion cannot reproduce the observed lithium depletion (1/140). Although part of the lithium depletion may occur during the pre main-sequence phase, the main-sequence lithium depletion is likely to be very much larger than that induced by microscopic diffusion alone. Lithium diffuses at about the same rate as helium, so that its abundance would decrease by some $10\%$ only by diffusion. On the other hand, mixing can bring matter from the outer layers down to the place where lithium is destroyed by nuclear reactions, which is much more efficient. Moreover diffusion builds a steep helium gradient below the convective zone which leads to a spike in the sound velocity compared to the seismic sun in this region. These results suggest the presence of some mild mixing below the convective zone of the internal Sun.

Macroscopic motions are needed to bring up to the convection zone the nuclearly lithium-depleted matter and to smooth the helium-gradient at the top of the radiative interior. On the other hand we now know that beryllium is probably not depleted in the Sun and that the $^3$He/$^4$He ratio cannot have changed by more than 10\% during these last 3 Gyrs (Geiss and Gloeckler, 1998). This means that the mixing must not bring too much helium to the convective region and that the beryllium nuclearly-depleted region must not be connected to the convective region through the mixing.

\begin{figure}
\centerline{\includegraphics[width=8cm]{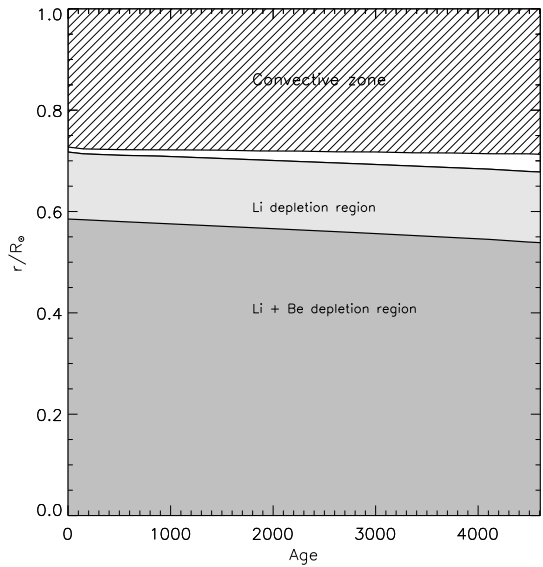}}
\caption{Location of lithium and beryllium nuclearly-depleted region with time (Myr) in a standard model including diffusion.}
\label{destruc}
\end{figure}

This put strong constraints on the mixing : it must be deep enough to deplete lithium and to smooth the helium-gradient below the convective region but it must be also shallow enough to keep constant the beryllium abundance and the $^3$He/$^4$He ratio during the last 3 Gyrs. 
We have represented on figure \ref{destruc}, the lithium and beryllium nuclearly-depleted regions as a function of the age in a standard model (including diffusion). According to the location of the lithium and beryllium nuclearly-depleted regions, the mixed zone may extend on a shallow region below the convective zone, down to 0.6$R$ at the beginning of the solar life (where $R$ represents the radius of the sun at that time) and down to 0.55$R_{\odot}$ at the solar age.

\subsection{Observational constraints on the tachocline}
Inversions of rotational splittings have shown that a strong horizontal
shear exists in the solar rotation profile at the base of the
convection zone in the so-called solar tachocline. Several attempts at
estimating the characteristic thickness of this region have been made from
helioseismic inversions. Values of this thickness range from $0.04-0.05
R_{\odot}$ (Basu, 1997; Antia et al., 1998; Basu and Antia, 2001; 
Charbonneau et al., 1998, 1999; Corbard et al., 1998) to $0.09R_{\odot}$ 
(Kosovichev, 1996; Kosovichev et al., 1998), the thickness being defined as the region where the
rotation rate variation inside the tachocline increases from 0.08 to 0.92 of its maximum value. More
recently Corbard et al. (1999) concluded that the
tachocline width is very likely below $0.05R_{\odot}$ and Elliot \& Gough (1999) 
obtained a value of $0.019 R_{\odot}$. 

As previously mentioned, the beryllium and $^3$He/$^4$He solar observations favor of a shallow mixed region below the convective region. Therefore the tachocline may provide an adequate mixing. However from helioseismic observations the tachocline is extremely thin and, supposing that its width has not been strongly increased in the past, the tachocline-induced mixing is not deep enough to produce a smooth helium gradient below the convective zone. Brun et al. (1999) have then considered a variable tachocline width during the solar evolution. In these conditions they have computed models in agreement with helioseismic and chemical observations. But there is no evidence and no explanation for a variable tachocline with time.

These results are in favor of extra mixing, probably related to the rotation-induced thermal imbalance. In the following we present solar models including both the effects of a tachocline and a rotation-induced mixing. In section \ref{mixing} we present the non-standard physical inputs introduced in the computations : we review the rotation-induced mixing theory together with the main equations, we discuss the stabilizing effects of $\mu$-gradients and we present the chosen tachocline modeling. Section \ref{computations} is devoted to the computational results. 

\section{Mixing theory}
\label{mixing}
\subsection{Rotation-induced mixing}
The theory of rotation-induced mixing in the presence of
$\mu$-gradients is introduced in the models. 

When the rotation depends little on latitude, the vertical component of the meridional velocity can be expressed in term of the Legendre function of order 2 :
\begin{equation}
u(r,\theta)=U_r(r)P_2(\cos \theta)
\end{equation}
$r$ being the radial coordinate and $\theta$ the colatitude. The horizontal component of the meridional velocity also depends on the vertical velocity amplitude $U_r(r)$ :
\begin{equation}
u_{\theta}=-\frac{1}{2 \rho r} \frac{d}{dr}(\rho r^2 U_r) \sin \theta \cos \theta
\end{equation}

In the case of the stationary regime and negligible differential rotation (as observed from helioseismic studies) the vertical amplitude can be written :
\begin{equation}
U_{r}=\frac{P}{\bar{\rho} \bar{g} \bar{T} C_p (\nabla_{ad}-\nabla+\nabla_{\mu})} \frac{L}{M_*}( E_{\Omega}+E_{\mu})
\end{equation}
$E_{\mu}$ gathers the terms related to the $\mu$-gradients while the classical terms appears in $E_{\Omega}$ :
\begin{equation}
E_{\Omega}=2 \left(1-\frac{\bar{\Omega}^2}{2 \pi G \bar{\rho}}\right) \frac{\tilde {g}}{\bar{g}}=\frac{8}{3} \frac{\Omega^2 r^3}{G M}\left(1-\frac{\bar{\Omega}^2}{2 \pi G \bar{\rho}}\right)  
\end{equation}
\begin{equation}
E_{\mu}=\frac{\rho_m}{\bar{\rho}}\left\{ \frac{r}{3} \frac{d}{dr} \left[ H_T
\frac{d\Lambda}{dr}  - \left(
\chi_{\mu}+\chi_T+1\right)
\Lambda \right] - 2\frac{H_T}{r}\Lambda\right\}
\end{equation}

Here $\displaystyle \overline \rho $ represents
the density average on the level surface 
$\displaystyle (\simeq \rho )$ while
$\displaystyle \rho_m$ is the mean density inside
the sphere of radius $\displaystyle r$; 
$\displaystyle H_T$ is the 
temperature scale height;
$\displaystyle \Lambda$  represents the 
horizontal $\displaystyle \mu $ fluctuations
$\displaystyle \frac{\tilde{\mu}}{\overline \mu} $;
$\displaystyle \chi _{\mu }$ and
$\displaystyle \chi _{T}$ represent the
derivatives:
\begin{equation}
\chi_{\mu } =
\left(
{\frac{\partial \ln \chi} { \partial \ln \mu } }\right)_{P,T}
\quad  ; \quad 
\chi_{T} =
\left( {\frac{\partial \ln \chi }{ \partial \ln T} }\right)_{P, \mu }
\end{equation}

Traditionally the meridional currents induced by the $E_{\Omega}$ term are called $\Omega$-currents while the ones induced by $E_{\mu}$ are called $\mu$-currents. In the most general case, $\mu$-currents are opposite to $\Omega$-currents.

The rotational mixing is then introduced in the computations following Zahn(1992) and Maeder \& Zahn (1998). The meridional circulation is supposed to induce shear flow instabilities which, in a density stratified medium may lead to large horizontal diffusivities. The combination of the meridional circulation and horizontal mixing may be treated as a vertical effective diffusion process described with an effective diffusion coefficient. The chemical transport for the mean concentration of a species $i$ is then expressed : 
\begin{equation}
\rho \frac{\partial{\bar{c_i}}}{\partial{t}}=  \frac{1}{r^2}
\frac{\partial}{\partial{r}} \biggl(r^2 \rho
D_{turb}\frac{\partial{\bar{c_i}}} {\partial{r}}\biggr)
\label{transport}
\end{equation}
where
\begin{equation}
D_{turb}=D_v+\frac{\left[r U_r(r)\right]^2}{30 D_h}
\end{equation}
$D_v$ and $D_h$ represent respectively the vertical and horizontal part of the shear-induced anisotropic turbulence due to the transport of angular momentum.

In the stationary case, according to Chaboyer \& Zahn (1992) :
\begin{equation}
\Lambda=- \frac{r^2}{6D_h} U_r \frac{\partial \ln \bar{\mu}}{\partial r}
\end{equation}

In the present paper, the horizontal diffusion coefficient is approximated following Maeder \& Zahn (1998) : 
\begin{equation}
D_h=C_h r|U_r|
\end{equation}
The efficiency of $D_v$ depends on the Richarson criterion (Maeder, 1995; Talon and Zahn, 1997). Here we chose to parametrize this coefficient in terms of $(r U_r)$ as for the vertical diffusion coefficient :
\begin{equation}
D_v=C_v r|U_r|
\end{equation}
In this case, $\Lambda$ can be written : 
\begin{equation}
\Lambda=- \frac{r}{6C_h} \frac{U_r}{|U_r|} \frac{\partial \ln \bar{\mu}}{\partial r}
\end{equation}
and we can also write :
\begin{equation}
D_{turb}= \biggl(C_v+\frac{1}{30C_h}\biggr) r|U_r| = \alpha_{turb} r|U_r|
\end{equation}
with $\displaystyle \alpha_{turb}=\biggl(C_v+\frac{1}{30C_h}\biggr)$. $C_v$ and $C_h$ are then unknown parameters. According to the assumption of strong anisotropic turbulence they must satisfied the condition :
\begin{equation}
C_v << C_h
\end{equation}

\subsection{Stabilizing effects of $\mu$-gradients}
It has been known for a long time that mixing processes in stars may
be stabilized in the regions where the mean molecular weight rapidly
decreases with increasing radius (Mestel, 1965; Huppert and Spiegel, 1977). 
This occurs specifically in the nuclear burning core. The importance of the feed-back effect due to nuclearly-induced $\mu$-gradients was well recognized (see Mestel and Moss, 1986), however the feed-back effect due to the diffusion-induced $\mu$-gradients on the meridional circulation was not included until recently. 
In
previous work on solar models, it was inferred that the rotation induced
mixing becomes inefficient as soon as the $\mu$-gradient becomes larger
than some critical value (RVCD). This value was
adjusted to obtain the lithium
depletion observed and to satisfy the constraint on the
${}^3$He/${}^4$He ratio.

Here we
include a more precise treatment of the physics of
diffusion-induced $\mu$-gradients and the diffusion-circulation coupling as
discussed by Vauclair \& Th\'eado (2003) (hereafter referred as paper I), Th\'eado \&
Vauclair (2003a, paper II) and (2003b, paper III). While the $\mu$-terms have definitely to be taken into account in the computations, the details of the coupling process which occurs when these term becomes of the same order of magnitude as the classical terms remain difficult to handle. In paper I, we introduced an analytical approach of the processes and discussed an approximate solution in a quasi-stationary case. We showed that the coupling between helium settling and rotation-induced mixing can slow down both the mixing and the settling when the $\mu$-terms are of the same order of magnitude as the other terms.
In paper II, we presented the results of a 2D numerical simulation of the considered processes which helps visualizing the situation. In the example which has been chosen (slowly rotating low mass stars), the $\mu$-terms become of the same order of magnitude as the classical terms in a short time scale compared to the main-sequence lifetime, thereby creating a frozen region where the circulation does not proceed anymore. However when this occurs the equilibrium between the opposite meridional currents is permanently destabilized by the helium settling below the convective zone. As a consequence, a new motion develops, which mixes up the zone polluted by diffusion. Below this region the circulation remains nearly frozen. This process acts as a restoring system for the element abundance variations and increases significantly their settling time scales. We then suggested that diffusion and mixing react in such a way as to keep both the horizontal and vertical $\mu$-gradients constant in the frozen region, while they proceed freely below.

\subsection{Tachocline}
\label{tac}
Macroscopic motions due to the tachocline can be introduced in the computations by adding a diffusion coefficient $D_{tacho}$ in the equation for the time
evolution of the concentration of chemical species. In this case the turbulent diffusion coefficient $D_{turb}$ which appears in equation \ref{transport} must be replaced by an effective diffusion coefficient $D_{eff}$ : 
\begin{equation}
D_{eff}=D_{turb}+D_{tacho}
\end{equation}

Spiegel \& Zahn (1992) and Brun et al. (1998b) have obtained an exponential diffusion coefficient for $D_{tacho}$. 
%Aspects of the shear layer are discussed in Brun et al. (1999).
Here the tachocline is treated as follows :
\begin{equation}
D_{tacho}=D_{cz} \exp (\ln 2 \frac{r-r_{cz}}{\Delta})
\label{dtacho}
\end{equation}
where $D_{cz}$ and $r_{cz}$ are respectively the value of $D_{tacho}$
and the value of the radius at the base of the
convective zone. $\Delta$ is called the half width of the tachocline. $D_{cz}$ and $\Delta$ are free parameters. 
Here the absolute size of the tachocline is assumed constant with time. 

\section{Computational results}
\label{computations}
The diffusion-circulation coupling and the tachocline described in section \ref{mixing} have been
introduced in the computations of solar models. 

\subsection{Constraining the free parameters}
\subsubsection{Rotation-induced mixing}
The free parameters of the rotation-induced mixing are $C_h$ and $C_v$ (or $\alpha_{turb}$) which determine the efficiency of the turbulent motions. The competition between mixing and diffusion then determine the construction of $\mu$-gradients and therefore also the strength of $\mu$-currents.

In particular the $\mu$-currents strongly depend on $C_h$ which is directly related to the horizontal turbulence : a strong horizontal turbulence tends to homogenize the horizontal layers and thereby smooths the horizontal $\mu$-gradients. In this case the small induced $\mu$-currents may not be able to compensate for the $\Omega$-currents and the mixing may remain very efficient during all the solar life. On the contrary, weak horizontal turbulence leads rapidly to important $\mu$-currents which may strongly reduce the mixing. 

In the computations, we try to adjust the mixing free parameters to produce a mixing both :
\begin{itemize}
\item efficient and deep enough to smooth the diffusion-induced helium gradient below the convective zone,
\item weak and shallow enough to prevent beryllium from nuclear destruction in the external layers of the Sun. 
\end{itemize}

\subsubsection{Tachocline}
By analogy with the usual definition of the tachocline (cf section \ref{tac}) we define the tachocline thickness $\omega$ as the region where the effective diffusion coefficient $D_{tacho}$ (cf eq. \ref{dtacho}) increases from 0.08 to 0.92 of its maximum value $D_{cz}$. As a result the constraint on $\omega$ deduced from helioseismic observations ($\omega<0.04R_{\odot}$) leads to a constraint on $\Delta$ which is :
\begin{equation}
\Delta < 0.01135 R_{\odot}
\end{equation}
Considering this constraint, the two free parameters of the tachocline ($\Delta$ and $D_{cz}$) are adjusted to reproduce the observed solar lithium depletion without depleting beryllium. For that we will assume that lithium depletion during the pre-main-sequence is negligible. As a consequence, the deduced values for the tachocline parameters will be upper limits since lithium may be partly depleted on pre-main-sequence.

\subsection{Results}

The characteristics of our best model are given in table \ref{soltab1}. This model is calibrated to reproduce at the solar age the observed luminosity and radius of the Sun. This calibration is achieved by adjusting the two main free parameters of the code which are the mixing length parameter $\alpha$ and the initial helium abundance $Y_0$. The obtained values are : $$\alpha=1.753739 \hspace{0.5cm} \rm{and} \hspace{0.5cm} {\it Y}_0=0.268039$$
\begin{table}
\caption{Constraints on solar models and characteristics of our best model.}
\medskip
\begin{tabular}{ccc}
\hline\noalign{\smallskip}
 & Constraints& Model \\
\hline\noalign{\smallskip}
Age (Gyrs)& 4.57$\pm$0.02  & 4.606 \\
M (g)&(1.9891$\pm$0.0004) 10$^{33}$ &1.9890 x 10$^{33}$\\
L (erg s$^{-1}$)& (3.8515$\pm$0.0055) 10$^{33}$  &3.8523 x 10$^{33}$ \\
R (cm)& (6.95749$\pm$0.00241) 10$^{10}$& 6.95654 x 10$^{10}$ \\
\hline\noalign{\smallskip}
r$_{\rm{bzc}}$/$R_{\odot}$& 0.713$\pm$0.003&0.718\\
$Y_{\rm{surf}}$ (EOS OPAL) & 0.249$\pm$0.002& 0.2507\\
$(Z/X)_{\odot}$ & 0.0245 & 0.0245\\
\hline\noalign{\smallskip}
$\rm{Li/Li_0}$  & 1/140 &1/133.8\\
$\rm{Be/Be_0}$ & 1 & 1/1.219\\
\hline\noalign{\smallskip}
$^3$He/$^4$He & $^3$He/$^4$He (t$_1$ $\rightsquigarrow$ t$_{\odot}$) $\le$ 10\%&\\
 \small{t$_1$ = 1.6 Gyrs}&  & 1.614 x 10$^{-4}$ \\
 \small{t$_{\odot}$ = 4.6 Gyrs} &  & 1.642 x 10$^{-4}$ \\
\noalign{\smallskip}
\hline
\end{tabular}
\label{soltab1}
\end{table}

As in RVCD, the rotation is assumed uniform in the models, but the rotational velocity decreases with time according to a Skumanich (1972) law, i.e. $\displaystyle \Omega \propto t^{-\frac{1}{2}}$. The coefficient is adjusted for an initial surface velocity V$_{R_{\odot}}$=100 km s$^{-1}$ and a surface velocity at the age of the Sun of V$_{R_{\odot}}$=2 km s$^{-1}$. 
In our computations, the rotation-induced mixing and the tachocline parameters have been chosen to reproduce the smooth diffusion-induced helium-gradient below the convective region and the observed lithium depletion while keeping the beryllium abundance constant. Their values are consistent with those proposed by  Th\'eado and Vauclair (2003b) for solar type stars.

This model is computed with the following values of the rotation induced mixing parameters : $$C_h=7000 \hspace{0.5cm}\rm{and}\hspace{0.5cm} \alpha_{\it turb}=1$$ and the following tachocline : $$D_{cz}=2.5 \times 10^5 \hspace{0.5cm} \rm{and} \hspace{0.5cm}\Delta=0.048 \times 10^{10}=6.900 \times 10^{-3}\it{R}_{\odot}$$
Both the conditions $C_h >> C_v$ and $\Delta < 0.01135 R_{\odot}$ are satisfied.

\begin{figure}
\includegraphics[width=6cm]{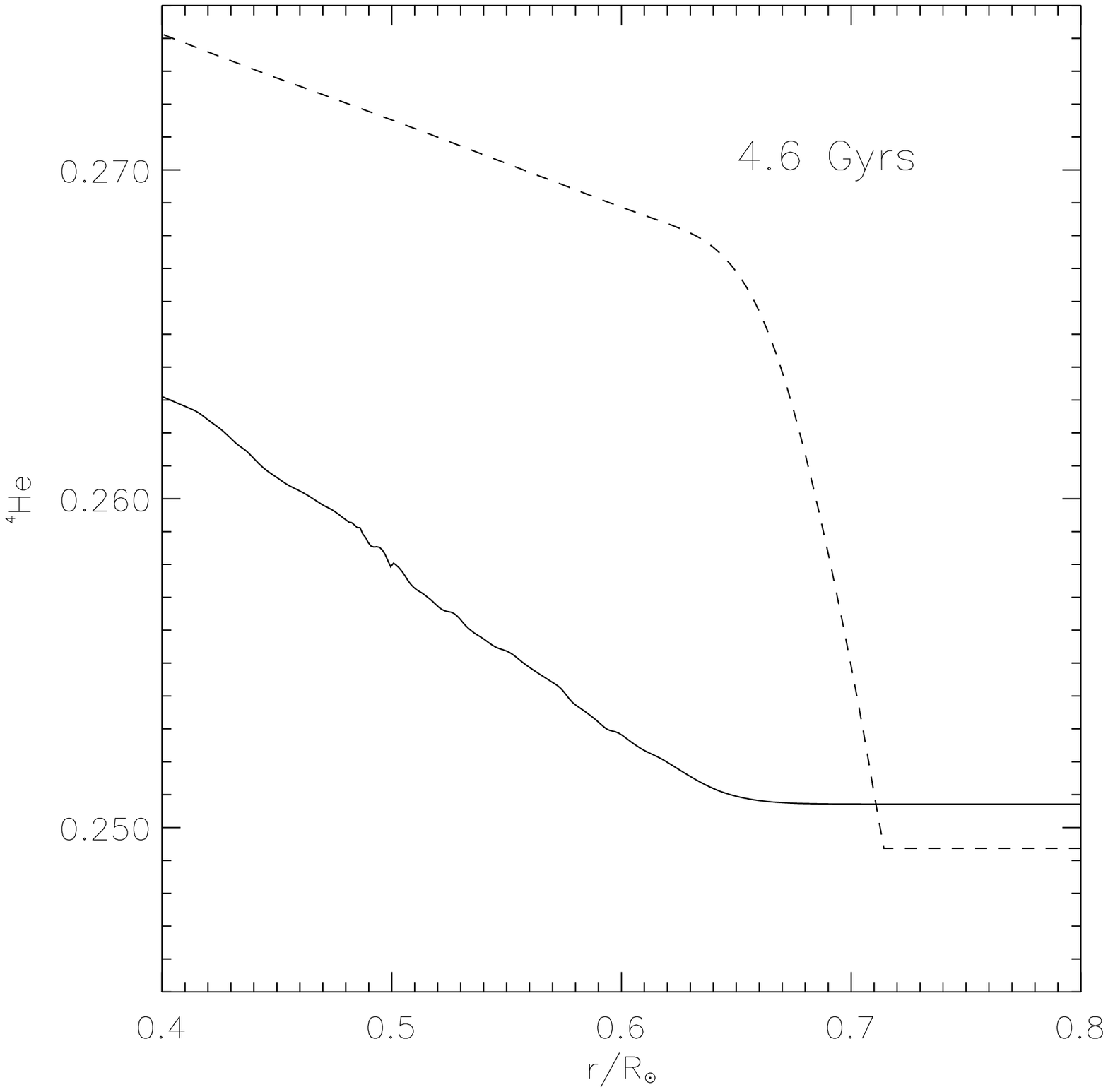}
\includegraphics[width=6cm]{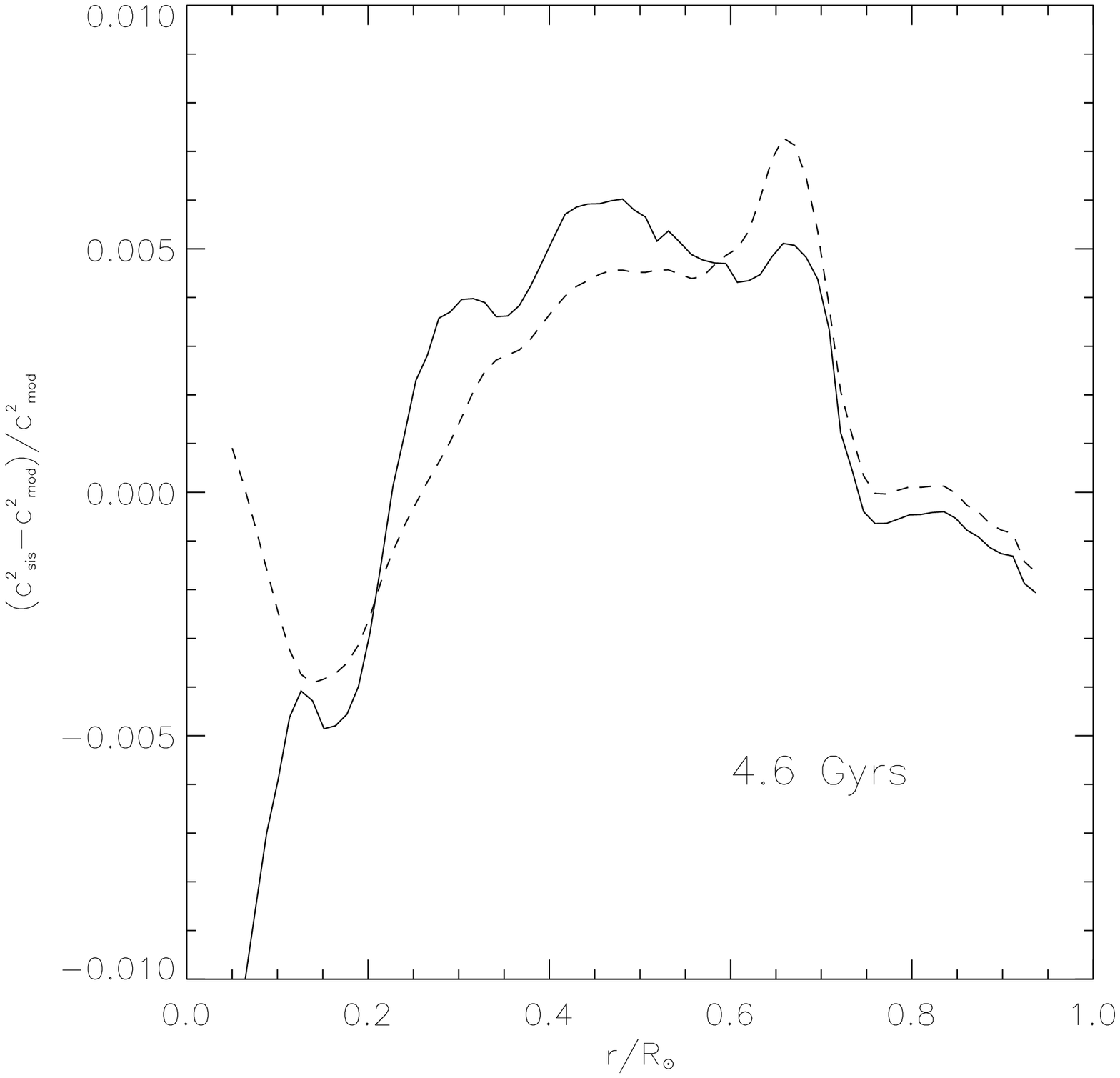}
\caption{Results at 4.606 Gyrs : the solid lines represent the results of our best model, the dashed lines show the results of a standard model including element segregation. Left panel : He profile below the convective zone. Right panel : comparison between $c^2$ (sound speed square) deduced from helioseismology (Basu et al., 1997) and the computed one.}
\label{ddc2}
\end{figure}

Figure \ref{ddc2} displays the helium gradient of our model below the convective region and it compares its sound speed square with the one deduced from helioseismic observations (Basu et al., 1997). For comparison the figure also shows the results obtained for a standard model including element segregation. 
Thanks to the mixing the helium-gradient is strongly smoothed and as a consequence the spike in the sound velocity observed in the case of pure element settling is reduced. However in the deep interior the rotation-induced mixing does not improve the agreement between models and helioseismology.

\begin{figure}
\centerline{\includegraphics[width=9cm]{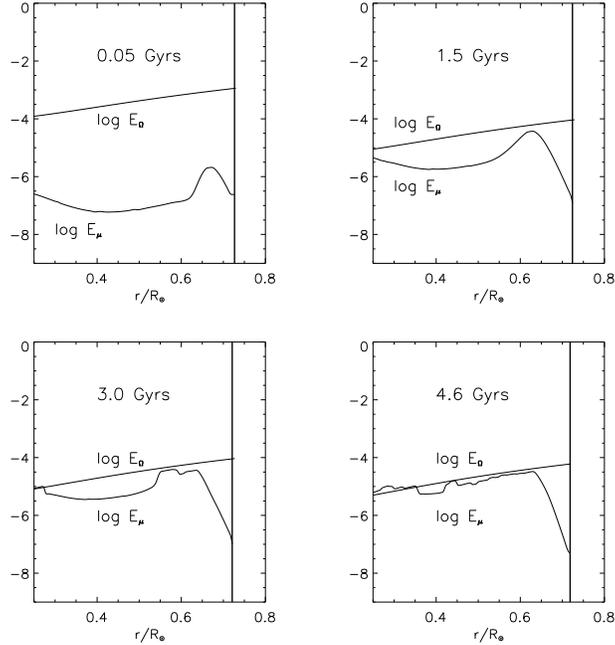}}
\caption{Evolution of $\Omega$ and $\mu$-currents with time inside the Sun. The graphs show the variations with depth of both $|E_{\Omega}|$ and $|E_{\mu}|$.}
\label{eomega}
\end{figure}

Figure \ref{eomega} shows the $|E_{\Omega}|$ and $|E_{\mu}|$ profiles inside the star at different evolutionary stages. The sun
arrives on the main-sequence with nearly homogeneous composition. At that
time $|E_{\mu}|$ is much smaller than $|E_{\Omega}|$ everywhere inside the
star. The meridional circulation and element settling can take place
in the radiative region of the star. Just below the convective zone, under the effects of the tachocline material is strongly mixed up and homogenized with the convective envelope. As a result in this region, the $\mu$-gradients cannot form and the $\mu$-currents remain negligible. On the other hand below the tachocline region, as a
consequence of the helium settling and the $\Omega$-currents, the $\mu$-currents increase, thereby reducing the mixing efficiency in a
non-negligible way as described in the previous section. They also increase in the core because of nuclear burning but other physical processes can
occur there which are not taken into account in the present computations (e.g. mild mixing in the internal layers at some stages of the solar evolution).  

\begin{figure}
\centerline{\includegraphics[width=9cm]{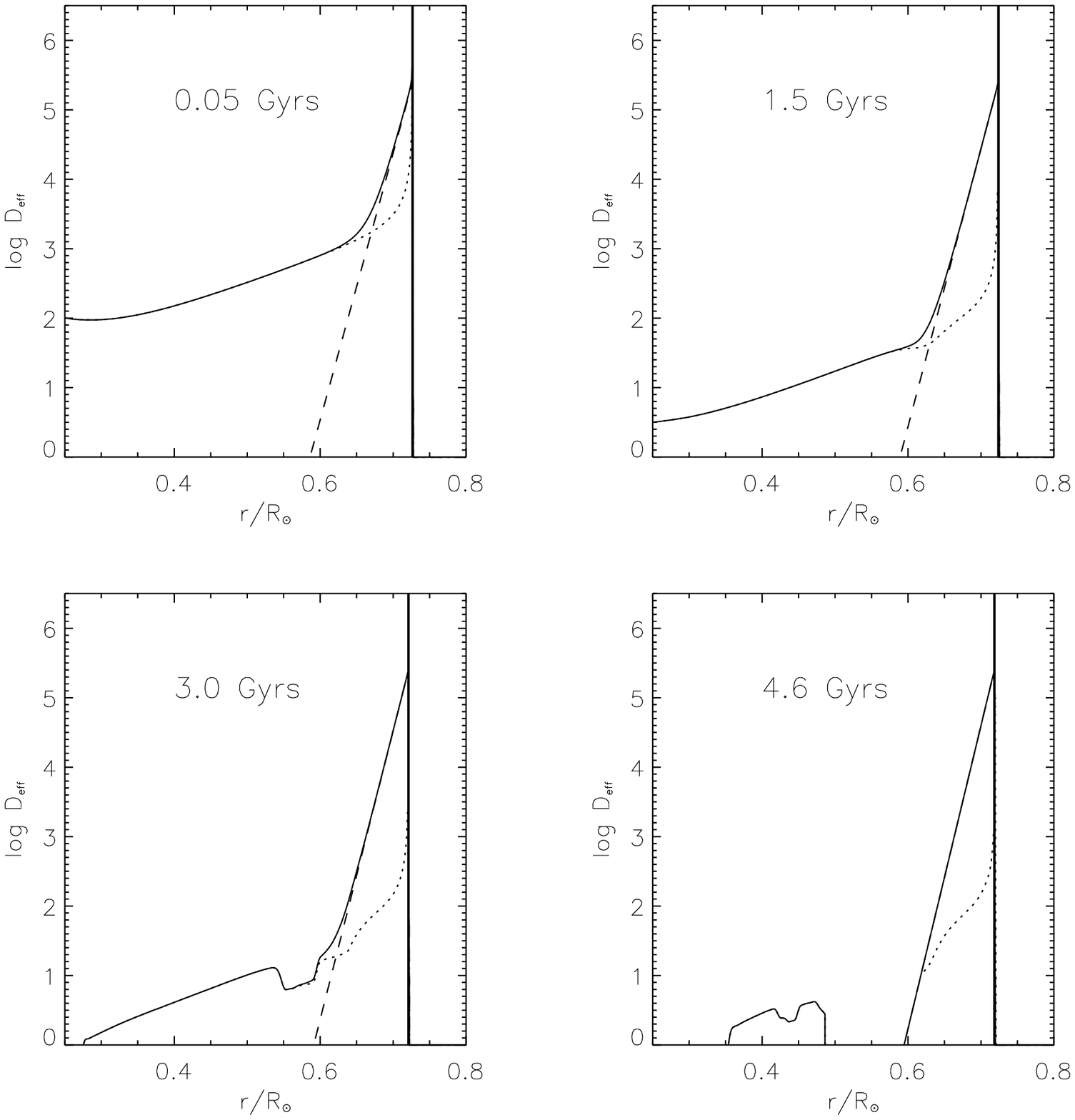}}
\caption{Evolution of the diffusion coefficients with time inside the Sun. The dotted lines show the turbulent diffusion coefficient $D_{turb}$, the dashed lines show the tachocline-induced diffusion coefficient $D_{tacho}$ and the solid lines show the effective diffusion coefficient $D_{eff}$=$D_{turb}$+$D_{tacho}$.}
\label{deff}
\end{figure}

Figure \ref{deff} displays the diffusion coefficients ($D_{eff}$, $D_{turb}$ and $D_{tacho}$) inside the star at different evolutionary stages. The tachocline-induced mixing is significant over a narrow region located above 0.6$R_{\odot}$ which means above the beryllium nuclear destruction region. Below the tachocline region, the mixing is due essentially to the rotation-induced mixing. In this region because of large $\mu$-currents, the diffusion coefficient becomes very small during the solar evolution.

Finally at the beginning of the solar life, a strong mixing takes place in the whole radiative interior. Just below the convective zone, in the tachocline region, this mixing remains very strong during all the solar evolution. Below the tachocline region, the meridional circulation become rapidly very weak (after 2 Gyrs). As a result the convective region is rapidly disconnected from the beryllium-nuclearly depleted region while it remains connected to the lithium one through the tachocline during all the evolution.
  
\begin{figure}
\includegraphics[width=6cm]{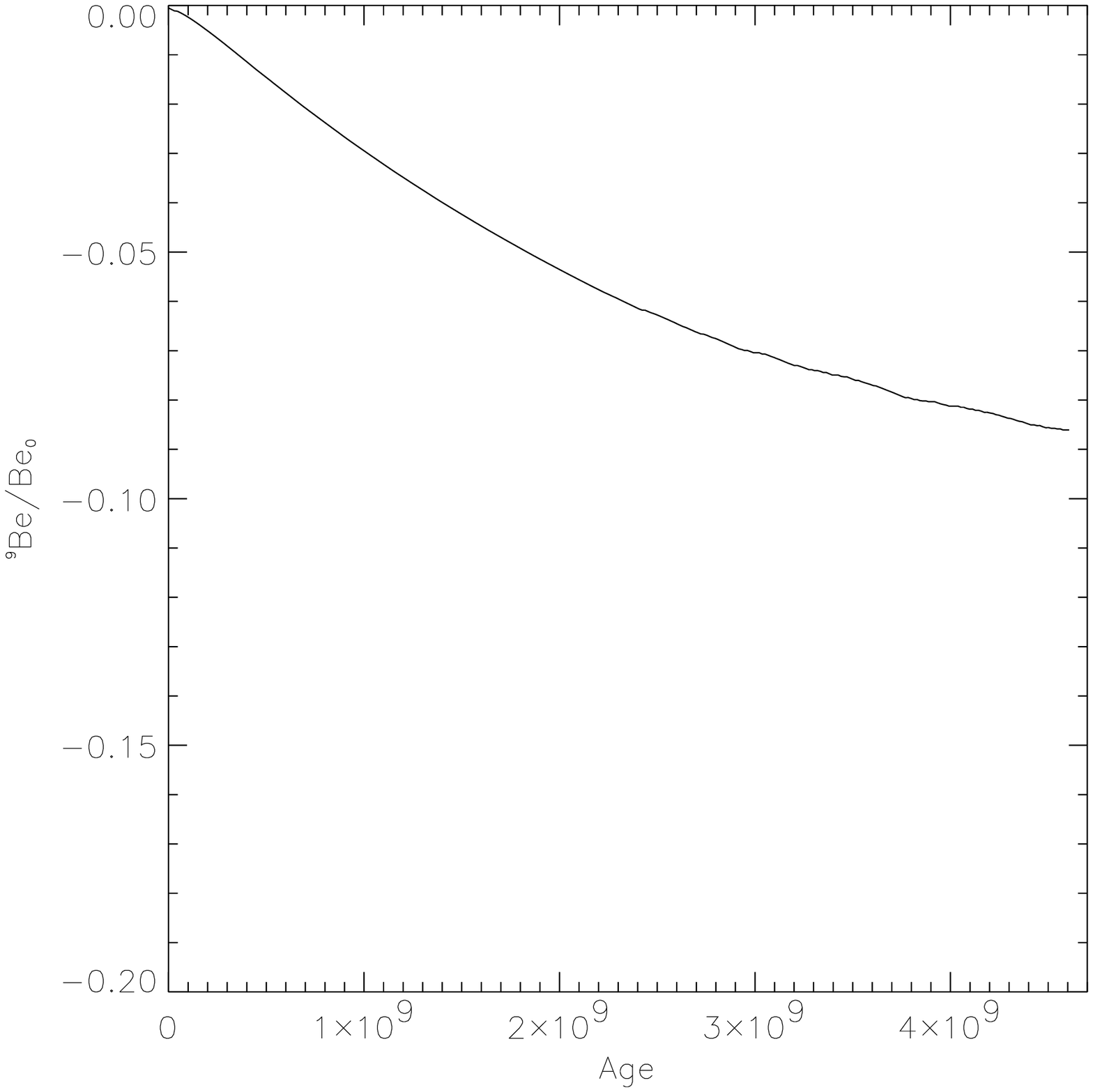}
\includegraphics[width=6cm]{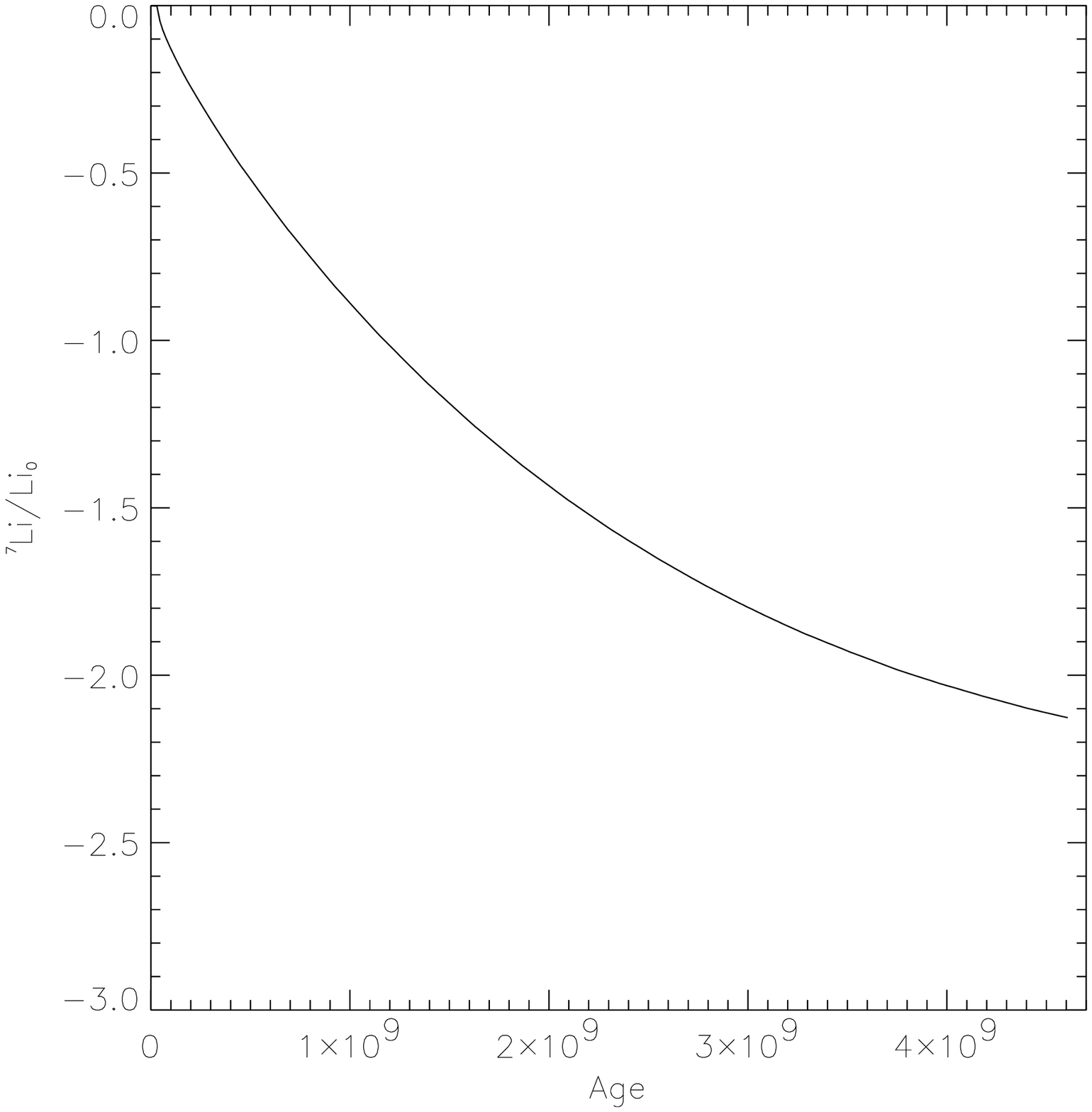}
\caption{Lithium and beryllium abundance variations with time (yrs) at the solar surface.}
\label{beli}
\end{figure}

Figure \ref{beli} displays the lithium and beryllium depletion with time at the solar surface. At the age of the Sun the lithium is depleted by a factor 133.8 while the beryllium depletion is lower than 18\%.

Therefore the chosen values of the free parameters reproduce quite well the lithium depletion observed in the Sun and lead to a small destruction of beryllium. Table \ref{soltab1} also shows that the constraints on the $^3$He/$^4$He ratio is satisfied as well as the constraints on the initial helium abundance ($Y_{surf}$=0.2507) which is
consistent with the value deduced from helioseismology by Basu \& Antia
(1997) and Richard et al. (1998) : $Y_{\odot,surf}=0.249 \pm 0.002$
(with the OPAL equation of state). The position of the convective region is however slightly overestimated.

The lithium abundance in the Sun may be compared to that observed in galactic clusters of known age. Table \ref{soltab2} then compares the lithium abundance in our model with that observed in four of these galactic clusters. The parametrization of the mixing and the tachocline below the convective zone correctly reproduces the lithium abundance evolution with time. 
\begin{table}
\caption{Lithium abundances observed in galactic clusters and lithium abundances of our model}
\medskip
\begin{tabular}{|c|cc|c|}
\hline%\noalign{\smallskip}
 & \multicolumn{2}{|c|}{Observations}& {Model}\\
Age & Cluster & A(Li) & A(Li)\\ 
\hline%\noalign{\smallskip}
0.05 Gyrs & $\alpha$ Persei & 3.2$\pm$0.3\,\footnote{}&3.25 \\

0.1 Gyrs&  Pleiades& 2.9$\pm$0.3\,\footnote{}&3.17\\

%0.5 Gyrs& Ursa Mayor &  &2.56 \\
0.8 Gyrs& Hyades&2.5$\pm$0.3\,\footnote{} & 2.55\\

1.7 Gyrs& NGC 752 &1.7$\pm$0.5\,\footnote{} &2.01\\
%\noalign{\smallskip}
\hline
\end{tabular}\\
%\begin{flushleft}
\footnotesize{$^1$ Balachandran et al. (1996)}
\footnotesize{$\hspace{1.3cm}$ $^3$ Thornburn et al. (1993)}\\
\footnotesize{$^2$ Soderblom et al. (1993)}
\footnotesize{$\hspace{2cm}$ $^4$ Balachandran (1995)}\\
%\end{flushleft}
\label{soltab2}
\end{table}

\section{Conclusion}
The chemical composition of the solar interior strongly relies on the competition between several physical processes : microscopic diffusion, rotation-induced mixing and tachocline mixing. Up to now, the feed-back effect of helium settling on the mixing was not taken into account in the computations, although the necessity for the mixing to stop at some depth below the convective zone appeared clearly from the observations of lithium, beryllium and helium 3. For this reason,  RVCD had introduced an artificial cut-off of the turbulent diffusion coefficient for a given $\mu$-gradient adjusted as a parameter. The present computations are more precise as they take into account the coupling between microscopic diffusion and rotation-induced mixing as described in previous papers by Th\'eado and Vauclair 2003 a and b. Mixing is naturally slowed down and stopped when the so-called $\mu$-currents are opposite to the $\Omega$-currents. The present modeling procedure also includes the most recent physical parameters for the opacities, equation of state, and nuclear reaction rates. The seismic sun is nicely reproduced by our best model, although some discrepancy by about 0.5 percent still remains. The same modeling procedure applied to solar-type stars in galactic clusters also reproduces correctly the lithium depletion with time.

\begin{acknowledgements}
We thank the referee, Joyce Guzik, for fruitful comments.
This work was supported by grant POCTI/1999/FIS/34549 approved by FCT
      and POCTI, with funds from the European Community programme FEDER
\end{acknowledgements}

\end{article}
\end{document}